\documentstyle[12pt,epsf,epsfig]{article}
\def\beq{\begin{equation}}
\def\eeq{\end{equation}}
\def\bea{\begin{eqnarray}}
\def\eea{\end{eqnarray}}
\def\bq{\begin{quote}}
\def\eq{\end{quote}}

\def\gappeq{\mathrel{\rlap {\raise.5ex\hbox{$>$}}
{\lower.5ex\hbox{$\sim$}}}}

\def\lappeq{\mathrel{\rlap{\raise.5ex\hbox{$<$}}
{\lower.5ex\hbox{$\sim$}}}}

\begin{document}
\begin{flushright}
{FISIST/23-2002/CFIF}\\
{SISSA 73/2002}
\end{flushright}
\vspace*{5mm}
\begin{center}
{\large {\bf MSSM with Yukawa Quasi-Unification}} \\
\vspace*{1cm}
{\bf M.~E.~G{\'o}mez$^1$ and C.~Pallis$^2$}\\ 
{\it
\vspace{0.3cm}
$^1$ Departamento de F\'{\i}sica and Grupo Te\'orico de F\'{\i}sica de
 Part\'{\i}culas, Instituto Superior T\'ecnico, \\
Av. Rovisco Pais, 1049-001~Lisboa, Portugal 

$^2$ Scuola Internazionale Superiore Di Studi
Avanzati (SISSA), \\
Via Beirut 2-4, 34013 Trieste, Italy}.\\
\vspace*{.5cm}
{\bf ABSTRACT} \\ 
\end{center}
%\vspace*{5mm}
\noindent
We consider the constrained minimal
supersymmetric standard model which emerges from
one theory with a small deviation from Yukawa
unification which is adequate for $\mu>0$. We show
that this model possesses a wide and natural range
of parameters which is consistent with the data on
$b\rightarrow s\gamma$, the muon anomalous magnetic
moment, the cold dark matter abundance in the
universe, and the Higgs boson masses.
%\vspace*{1cm}
\noindent

%\setcounter{page}{1}
%\pagestyle{empty}

%INSERT YOUR TEXT HERE

%%%%%%%%%%%%%%%%%%%%%%%%%%%%%%%%%%%%%%%%%%%%%%%%%%%%%%%%%%%%%%%%%%%%%
\section{Introduction}

We study the phenomenological consequences of imposing 
on the constrained MSSM (CMSSM) an asymptotic relation for the 
Yukawa couplings at the GUT scale. This assumption
(Yukawa unification) naturally restricts \cite{als} the
$t$-quark mass to large values compatible with the data.
Also, the emerging model is highly predictive \cite{copw}. Despite of 
its appealing, the 
simple scheme of a single Yukawa for the three third generations at 
the GUT scale leads to an unacceptable $b$--quark mass. This fact 
excludes minimal versions of GUT groups with this property, such 
as Pati--Salam unification $(G_{PS}=SU(4)_c\times SU(2)_L\times
SU(2)_R )$, $SO(10)$ or $E_6$. 

We consider the SUSY GUT model described in Ref.\cite{Gomez:2002tj}
which is based in the $G_{PS}$ as described in Refs.\cite{jean}
and  establish an `asymptotic' relation for the Yukawa couplings that 
depends on a single complex parameter $c$:
\begin{equation}
h_t:h_b:h_\tau=|1+c|:|1-c|:|1+3c|,
\label{minimal}
\end{equation}
 For simplicity, we will restrict our analysis to real
values of $0<c<1$. The relative splitting of the Yukawa 
couplings becomes: $\delta h\equiv-(h_b-h_t)/h_t=
(h_\tau-h_t)/h_t=2c/(1+c)$. This means that the bottom and tau Yukawa
couplings split from the top Yukawa coupling by
the same amount but in opposite directions, with
$h_b$ becoming smaller than $h_t$.

\section{The MSSM with Quasi-Yukawa Unification}

\label{mssm}
This model, below $M_{GUT}$, reduces to the MSSM supplemented by
the `asymptotic' Yukawa coupling quasi-unification
condition in Eq.(\ref{minimal}). We will assume 
universal soft
SUSY breaking terms at $M_{GUT}$, i.e., a common mass
for all scalar fields $m_0$, a common gaugino mass
$M_{1/2}$ and a common trilinear scalar coupling
$A_0$. In the present work, we will concentrate on the $\mu>0$.
The case  $\mu<0$ is phenomenologically less interesting, it will be 
presented in \cite{costas}. We follow the notation as well as the
RG and radiative electroweak breaking analysis of
Ref.\cite{cdm} for the CMSSM with the improvements
of Refs.\cite{cd2, Gomez:2002tj} (recall that the sign of
$\mu$ in Refs.\cite{cdm,cd2} is opposite to 
Ref.\cite{Gomez:2002tj}, which is the one 
adopted here). 

For any given $m_b(M_Z)$ in its $95\%$ c.l.
range ($2.684-3.092~{\rm GeV}$ for $\alpha_s(M_Z)=0.1185$), we 
can determine
the parameters $c$ and $\tan\beta$ at
$M_{SUSY}=(m_{\tilde t_1}
m_{\tilde t_2})^{1/2}$ ($\tilde t_{1,2}$ are the
stop mass eigenstates) 
so that the `asymptotic' condition in
Eq.(\ref{minimal}) is satisfied. We use fixed
values for the running top quark mass
$m_t(m_t)=166~{\rm GeV}$ and the running tau lepton
mass $m_\tau(M_Z)=1.746~{\rm{GeV}}$ and incorporate
not only the SUSY correction to the bottom quark
mass but also the SUSY threshold correction to
$m_\tau(M_{SUSY})$ from the approximate formula of
Ref.\cite{pierce}.  After imposing the conditions of gauge coupling
unification, successful electroweak breaking and
Yukawa quasi-unification in Eq.(\ref{minimal}),
we are left with three free input parameters $m_0$,
$M_{1/2}$ and $A_0$. In
order to make the notation physically more
transparent, we replace $m_0$ and $M_{1/2}$
equivalently by the mass $m_{LSP}$ (or
$m_{\tilde\chi}$) of the lightest supersymmetric
particle (LSP), which turns out to be the lightest
neutralino ($\tilde\chi$), and the relative mass
splitting  $\Delta_{\tilde\tau_2}=(m_{\tilde\tau_2}
-m_{\tilde\chi})/m_{\tilde\chi}$ between the lightest
stau mass eigenstate ($\tilde\tau_2$) and the LSP. In 
Fig.1 we display the changes on $M_{SUSY}$ and the mass of the 
pseudo-scalar Higgs, $m_A$, for several values of 
$\Delta_{\tilde\tau_2}$, and $m_b(M_Z)$. These changes 
will help us to understand the corresponding predictions 
for $\Omega_{LSP} h^2$ in the presence of 
resonant annihilation channels for values 
of $m_A\approx 2\cdot m_{LSP}$.
\begin{figure}[ht]
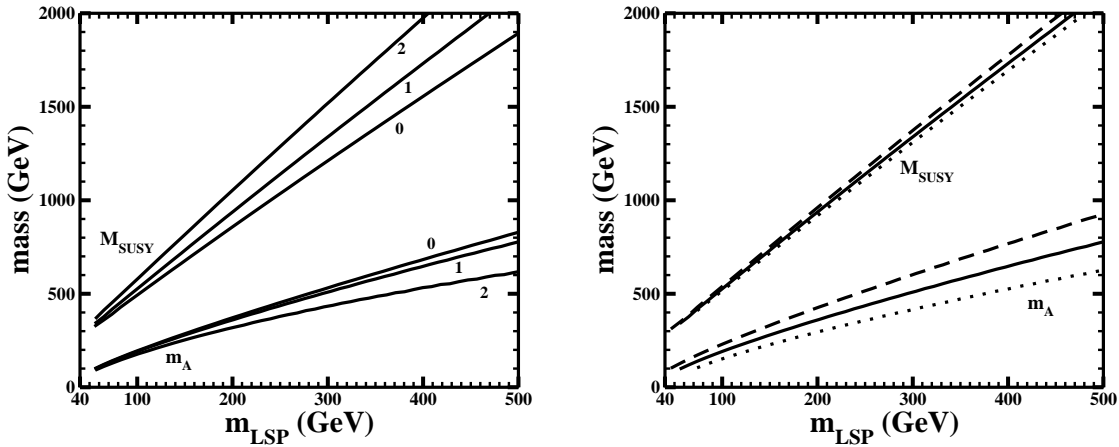

\hspace*{-.35in}
\begin{center}
\begin{minipage}{6in}

\epsfig{figure=massfig.eps,height=2.3in,angle=0}
\hspace*{0.5 cm}
\epsfig{figure=mbmassfig.eps,height=2.3in,angle=0}
\hfill
\end{minipage} 
\end{center}
\hfill
\caption[]{\it
\footnotesize
%\par {FIG. \Figb.} 
The mass parameters $m_A$ and
$M_{SUSY}$ as functions of $m_{LSP}$, with  $\alpha_s(M_Z)=0.1185$ 
and $A_0=0$. 
On the panel of the left we display various
values of $\Delta_{\tilde\tau_2}$ (indicated on the curves) for 
$m_b(M_Z)=2.888~{\rm GeV}$.
On the right panel we fix $\Delta_{\tilde\tau_2}=1$
and show the curves for 
$m_b(M_Z)=2.684~{\rm GeV}$ (dashed lines),
$3.092~{\rm GeV}$ (dotted lines) or
$2.888~{\rm GeV}$ (solid lines).}
\end{figure}

Here, the LSP ($\tilde\chi$) is an almost pure bino.
Its relic abundance will be calculated by
{\tt micrOMEGAs} \cite{micro}, which is the most
complete code available. It includes all the
coannihilations \cite{coan} of neutralinos,
charginos, sleptons, squarks and gluinos since 
it incorporates automatically all possible 
channels by using {\tt COMPHEP} \cite{Pukhov} 
(A similar calculation has
appeared in Ref.\cite{baer}.) Also, poles
and thresholds are properly handled and one-loop
QCD corrected Higgs decay widths \cite{width}
are used, which is the main improvement
provided by Ref.\cite{micro}. The SUSY
corrections \cite{susy} to these widths are,
however, not included. Fortunately, in our case,
their effect is much smaller than that of the
QCD corrections. From the
recent results of DASI \cite{dasi}, one finds that the
$95\%$ c.l. range of $\Omega_{CDM}~h^{2}$ is $0.06-0.22$.
Therefore, we require that $\Omega_{LSP}~h^2$ does
not exceed 0.22.

\begin{figure}[tb]
\begin{center}
\epsfig{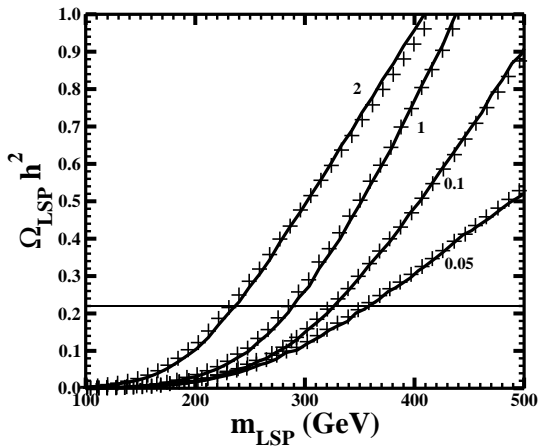}
\end{center}
\caption[]{\it 
\footnotesize
%\par {FIG. \Figc.} 
The LSP relic abundance
$\Omega_{LSP}~h^2$ versus $m_{LSP}$ for various
$\Delta_{\tilde\tau_2}$'s (indicated on the curves)
and with $A_0=0$, $m_b(M_Z)=2.888~{\rm GeV}$,
$\alpha_s(M_Z)=0.1185$. The solid lines (crosses)
are obtained by {\tt micrOMEGAs} (our alternative
method). The upper bound on $\Omega_{LSP}~h^2$
(=0.22) is also depicted.}
\end{figure}

In order to have an independent check of
{\tt micrOMEGAs}, we also use the following alternative
method for calculating $\Omega_{LSP}~h^2$ in
our model. In most of the parameter space where
coannihilations are unimportant, $\Omega_{LSP}~h^2$
can be calculated by using {\tt DarkSUSY}
\cite{darksusy} \footnote{ An updated version of this code is now available 
\cite{Edsjo:2003us}.}. Its neutralino annihilation part is in excellent
numerical agreement with the recent exact analytic
calculation of Ref.\cite{roberto}, the main defect of its current 
version  
is that it uses the tree-level
Higgs decay widths. This can be approximately corrected
if, in evaluating the Higgs decay widths, we replace
$m_b(m_b)$ by $m_b$ at the mass of the appropriate
Higgs boson in the couplings of the $b$-quark to the
Higgs bosons (see Ref.\cite{micro}). In the region of the parameter 
space where coannihilations come into play, the next-to-lightest
supersymmetric particle (NLSP) turns out to be the
$\tilde\tau_2$ and the only relevant coannihilations
are the bino-stau ones \cite{cdm,ellis} (we do not find bino-stop 
cannihilations  \cite{djouadi}, potencially important when $A$ is negative). 
In this
region, which is given by
$\Delta_{\tilde\tau_2}<0.25$, we calculate
$\Omega_{LSP}~h^2$ by using an improved version
of the analysis of Ref.\cite{cdm,cd2,hw,vergados}. The  list of 
bino-stau coannihilation 
channels appropriate for all $\tan\beta$'s  given  
Ref.~\cite{cdm} has been completed with some additional channels
as described in \cite{Gomez:2002tj}(see also
Refs.\cite{ellis,arnowitt}). Their corresponding cross sections 
are combined with the results of {\tt DarkSUSY} as described 
in \cite{Gomez:2002tj}. The results presented  in Fig.~2 show an  
impressive agreement of the two methods.

We calculate ${\rm BR}(b\rightarrow s\gamma)$ using the
formalism of Ref.\cite{kagan}, where the SM contribution
is factorized out. This contribution includes the
next-to-leading order (NLO) QCD and the leading order (LO)
QED corrections. The charged Higgs boson contribution to
${\rm BR}(b\rightarrow s\gamma)$ is evaluated by including
the NLO QCD corrections from Ref.\cite{nlohiggs}. The
dominant SUSY contribution includes the NLO QCD
corrections from Ref.\cite{nlosusy}, which hold for large
$\tan\beta$. With the considerations
of \cite{Gomez:2002tj} the $95\%$ c.l.
range of this branching ratio then turns out to be about
$(1.9-4.6)\times 10^{-4}$. 

According with the latest measurement \cite{muon} of the 
anomalous magnetic moment of the muon $a_\mu\equiv (g_\mu-2)/2$,  
the deviation of from its predicted value in the SM  \cite{davier}, 
$\delta a_\mu$, is
found to lie, at $95\%$ c.l., in the range from $-4.7\times
10^{-10}$ to $56\times 10^{-10}$ when the SM calculations based in 
$e^+ e^-$ data and in $\tau$ data are both taken into account. The
calculation of $\delta a_\mu$ in the CMSSM is performed
here by using the analysis of Ref.\cite{gminus2}, the updating 
of the experimental bounds does not introduce significant 
differences respect the results presented in \cite{Gomez:2002tj}.

We will also impose the $95\%$ c.l. LEP bound on the
lightest CP-even neutral Higgs boson mass
$m_h>114.1~{\rm GeV}$. In the CMSSM, this
bound holds almost always for all $\tan\beta$'s, at least
as long as CP is conserved. The CP-even neutral Higgs boson
mass matrix by using {\tt FeynHiggsFast} \cite{fh}. Finally, 
for the values of
$\tan\beta$ which appear here (about 60), the CDF results
yield the $95\%$ c.l. bound $m_A>110~{\rm GeV}$ \cite{cdf}.

\begin{figure}[tb]
\begin{center}
\epsfig{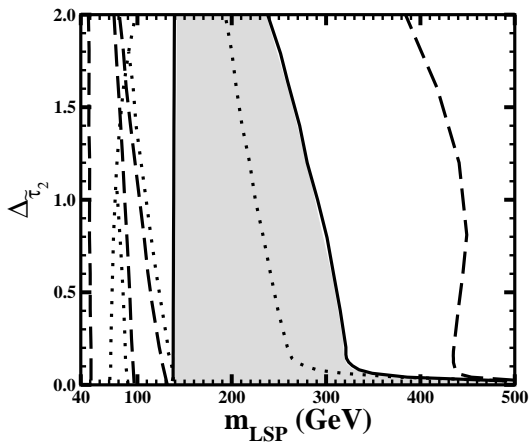}
\end{center}
\caption[]{\it \footnotesize
%\par {FIG. \Figa.} 
Restrictions on the
$m_{LSP}-\Delta_{\tilde\tau_2}$ plane for $A_0=0$,
$\alpha_s(M_Z)=0.1185$. From left to right, the
dashed (dotted) lines depict the lower bounds on
$m_{LSP}$ from $m_A>110~{\rm GeV}$,
${\rm BR}(b\rightarrow s\gamma)>1.9\times 10^{-4}$
and $\delta a_\mu<58\times 10^{-10}$, and the
upper bound on $m_{LSP}$ from
$\Omega_{LSP}~h^2<0.22$ for
$m_b(M_Z)=2.684~{\rm GeV}$
($3.092~{\rm GeV}$). The left (right) solid
line depicts the lower (upper) bound on $m_{LSP}$
from $m_h>114.1~{\rm GeV}$ ($\Omega_{LSP}~h^2<0.22$)
for $m_b(M_Z)=2.888~{\rm GeV}$. The allowed area for
$m_b(M_Z)=2.888~{\rm GeV}$ is shaded.}
\end{figure}

\section{The Allowed Parameter Space}
\label{parameters}
\par
\vspace{-.5cm}
The
restrictions on the $m_{LSP}-\Delta_{\tilde\tau_2}$
plane, for $A_0=0$ and with the central value of
$\alpha_s(M_Z)=0.1185$, are shown in Fig.3. The
dashed (dotted) lines correspond to the $95\%$ c.l.
lower (upper) experimental bound on $m_b(M_Z)$ which
is $2.684~{\rm GeV}$ ($3.092~{\rm GeV}$), while the solid lines
correspond to the central experimental value of
$m_b(M_Z)=2.888~{\rm GeV}$. From
left to right, the dashed (dotted) lines depict the lower 
bounds on $m_{LSP}$ from the
constraints $m_A>110~{\rm GeV}$,
${\rm BR}(b\rightarrow s\gamma)>1.9\times 10^{-4}$
and $\delta a_\mu<58\times 10^{-10}$, and the $95\%$
c.l. upper bound on $m_{LSP}$ from
$\Omega_{LSP}~h^2<0.22$. The constraints
${\rm BR}(b\rightarrow s\gamma)<4.6\times 10^{-4}$
and $\delta a_\mu>-6\times 10^{-10}$ do not restrict
the parameters since they are always satisfied for
$\mu>0$. The left solid line depicts the lower bound
on $m_{LSP}$ from $m_h>114.1~{\rm GeV}$ which does not
depend much on $m_b(M_Z)$, while the right solid line
corresponds to $\Omega_{LSP}~h^2=0.22$ for the central
value of $m_b(M_Z)$. We see that $m_A$ is always smaller than
$2m_{LSP}$ but close to it. Thus, generally, the
neutralino annihilation via the s-channel exchange
of an $A$-boson is by far the dominant
(co)annihilation process. We also observe that,
as $m_{LSP}$ or $\Delta_{\tilde\tau_2}$ increase,
we move away from the $A$-pole, which thus becomes
less efficient. As a consequence,
$\Omega_{LSP}~h^2$ increases with $m_{LSP}$ or
$\Delta_{\tilde\tau_2}$ (see Fig.2).

In the allowed (shaded) area of Fig.~3 which
corresponds to the central value of $m_b(M_Z)$,
the parameter $c$ ($\tan\beta$) varies between
about 0.15 and 0.20 (58 and 59). For $m_b(M_Z)$
fixed to its lower or upper bound, we find that,
in the corresponding allowed area, the parameter
$c$ ($\tan\beta$) ranges between about 0.17 and
0.23 (59 and 61) or 0.13 and 0.17 (56 and 58).
We observe that, as we increase $m_b(M_Z)$, the
parameter $c$ decreases and we get closer to
exact Yukawa unification. This behavior is
certainly consistent with the fact that the
value of $m_b(M_Z)$ which corresponds to exact
Yukawa unification lies well above its $95\%$
c.l. range. The LSP mass is restricted to be
higher than about $138~{\rm GeV}$ for $A_0=0$
and $\alpha_s(M_Z)=0.1185$, with the minimum being
practically $\Delta_{\tilde\tau_2}$-independent.
At this minimum, $c\approx 0.16-0.20$
($c\approx 0.13-0.23$) and
$\tan\beta\approx 59$
($\tan\beta\approx 58-61$) for
$m_b(M_Z)=2.888~{\rm GeV}$
($m_b(M_Z)=2.684-3.092~{\rm GeV}$).

\begin{figure}[ht]
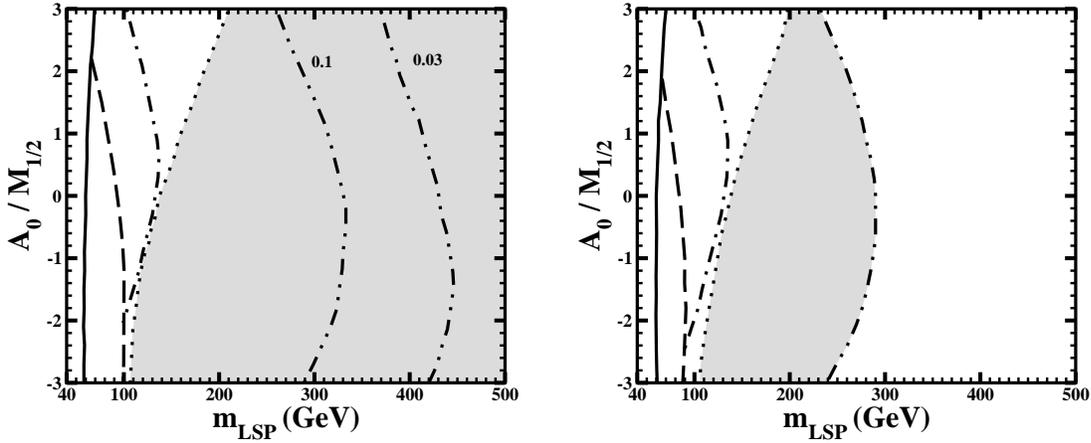

\hspace*{-.35in}
\begin{center}
\begin{minipage}{6in}

\epsfig{figure=D0.eps,height=2.3in,angle=0}
\hspace*{0.5 cm}
\epsfig{figure=D1.eps,height=2.3in,angle=0}
\hfill
\end{minipage} 
\end{center}
\hfill
\vspace{-.5cm}
\caption[]{\it
\footnotesize Restrictions on the
$m_{LSP}-A_0/M_{1/2}$ plane for
$\Delta_{\tilde\tau_2}=0$ (left) and 1 (right), 
$m_b(M_Z)=2.888~{\rm GeV}$,
$\alpha_s(M_Z)=0.1185$. The solid, dashed,
dot-dashed and dotted lines correspond to the lower
bounds on $m_{LSP}$ from $m_A>110~{\rm GeV}$,
${\rm BR}(b\rightarrow s\gamma)>1.9\times 10^{-4}$,
$\delta a_\mu<58\times 10^{-10}$ and
$m_h>114.1~{\rm GeV}$ respectively. The upper bound
on $m_{LSP}$ from $\Omega_{LSP}~h^2<0.22$ does not
appear in the left panel since it lies at
$m_{LSP}>500~{\rm GeV}$. The allowed area is shaded.
For comparison, we also display on the left panel the bounds from
$\Omega_{LSP}~h^2<0.22$ (double dot-dashed lines)
for $\Delta_{\tilde\tau_2}=0.1~{\rm and}~0.03$, as
indicated. The upper bound on $m_{LSP}$ from the cosmological
constraint $\Omega_{LSP}~h^2<0.22$ corresponds to the double 
dot-dashed line on the left panel.}
\end{figure}

In Fig.~3, we present the restrictions on 
the $m_{LSP}-A_0/M_{1/2}$ plane for $m_b(M_Z)=2.888~{\rm GeV}$,
$\alpha_s(M_Z)=0.1185$ and fixed values of $\Delta_{\tilde\tau_2}$. 
The most significant restriction on the allowed area are 
due to the displacement of the $\Omega_{LSP}~h^2=0.22$ line as 
$\Delta_{\tilde\tau_2}$ increases, showing clearly the effect of the 
bino-stau coannihilations on the left panel. On the right panel we 
can observe that the allowed area becomes narrower as 
$|A_0/M_{1/2}|\neq 0$.
\vspace{-.5cm}
\section{Conclusions}
\label{conclusions}
\vspace{-.5cm}
\par
We showed that, in the particular model with Yukawa
quasi-unification considered, there exists a wide
and natural range of CMSSM parameters which is
consistent with all the above constraints. We found
that, within the investigated part of the overall
allowed parameter space, the parameter $\tan\beta$
ranges between about 58 and 61 and the `asymptotic'
splitting between the bottom (or tau) and the top
Yukawa couplings varies in the range $26-35\%$ for
central values of $m_b(M_Z)$ and $\alpha_s(M_Z)$.

{\bf Acknowledgements.}
This work was supported by European
Union under the RTN contracts HPRN-CT-2000-00148 and
HPRN-CT-2000-00152. M.E.G. acknowledges support from
the `Funda\c c\~ao para a Ci\^encia e Tecnologia'
under contract SFRH/BPD/5711/2001 and project CFIF-Plurianual (2/91).

\def\ijmp#1#2#3{{Int. Jour. Mod. Phys. }{\bf #1~}(#2)~#3}
\def\plb#1#2#3{{Phys. Lett. }{\bf B~#1~}(#2)~#3}
\def\zpc#1#2#3{{Z. Phys. }{\bf C~#1~}(#2)~#3}
\def\prl#1#2#3{{Phys. Rev. Lett. }{\bf #1~}(#2)~#3}
\def\rmp#1#2#3{{Rev. Mod. Phys. }{\bf #1~}(#2)~#3}
\def\prep#1#2#3{{Phys. Rep. }{\bf #1~}(#2)~#3}
\def\prd#1#2#3{{Phys. Rev. }{\bf D~#1~}(#2)~#3}
\def\npb#1#2#3{{Nucl. Phys. }{\bf B~#1~}(#2)~#3}
\def\npps#1#2#3{{Nucl. Phys. (Proc. Sup.) }
{\bf B~#1~}(#2)~#3}
\def\mpl#1#2#3{{Mod. Phys. Lett. }{\bf #1~}(#2)~#3}
\def\arnps#1#2#3{{Annu. Rev. Nucl. Part. Sci. }
{\bf #1~}(#2)~#3}
\def\sjnp#1#2#3{{Sov. J. Nucl. Phys. }{\bf #1~}(#2)~#3}
\def\jetp#1#2#3{{JETP Lett. }{\bf #1~}(#2)~#3}
\def\app#1#2#3{{Acta Phys. Polon. }{\bf #1~}(#2)~#3}
\def\rnc#1#2#3{{Riv. Nuovo Cim. }{\bf #1~}(#2)~#3}
\def\ap#1#2#3{{Ann. Phys. }{\bf #1~}(#2)~#3}
\def\ptp#1#2#3{{Prog. Theor. Phys. }{\bf #1~}(#2)~#3}
\def\apjl#1#2#3{{Astrophys. J. Lett. }{\bf #1~}(#2)~#3}
\def\n#1#2#3{{Nature }{\bf #1~}(#2)~#3}
\def\apj#1#2#3{{Astrophys. Journal }{\bf #1~}(#2)~#3}
\def\anj#1#2#3{{Astron. J. }{\bf #1~}(#2)~#3}
\def\mnras#1#2#3{{MNRAS }{\bf #1~}(#2)~#3}
\def\grg#1#2#3{{Gen. Rel. Grav. }{\bf #1~}(#2)~#3}
\def\s#1#2#3{{Science }{\bf #1~}(#2)~#3}
\def\baas#1#2#3{{Bull. Am. Astron. Soc. }{\bf #1~}(#2)~#3}
\def\ibid#1#2#3{{ibid. }{\bf #1~}(#2)~#3}
\def\cpc#1#2#3{{Comput. Phys. Commun. }{\bf #1~}(#2)~#3}
\def\astp#1#2#3{{Astropart. Phys. }{\bf #1~}(#2)~#3}
\def\epjc#1#2#3{{Eur. Phys. J. }{\bf C~#1~}(#2)~#3}
\def\nima#1#2#3{{Nucl. Instrum. Meth. }{\bf A~#1~}(#2)~#3}
\def\jhep#1#2#3{{JHEP }{\bf #1~}(#2)~#3}

\end{document}